\documentclass[aps,reprint,pra,amstext,amsmath,superscriptaddress,floatfix]{revtex4-2}

\usepackage{graphicx}
\usepackage{xspace}
\usepackage{dcolumn}
\usepackage{bm}
\usepackage{xcolor}

\begin{document}
\title{Possible unconventional order parameter in single crystals of SrPt$_{3}$P
superconductor}
\author{Kyuil~Cho}
\affiliation{Ames Laboratory, Ames, Iowa 50011, USA}
\email[Corresponding author: ]{cho@hope.edu}
\author{S.~Teknowijoyo}
\affiliation{Ames Laboratory, Ames, Iowa 50011, USA}
\affiliation{Department of Physics \& Astronomy, Iowa State University, Ames, Iowa
50011, USA}
\author{S.~Ghimire}
\affiliation{Ames Laboratory, Ames, Iowa 50011, USA}
\affiliation{Department of Physics \& Astronomy, Iowa State University, Ames, Iowa
50011, USA}
\author{E.~H.~Krenkel}
\affiliation{Ames Laboratory, Ames, Iowa 50011, USA}
\affiliation{Department of Physics \& Astronomy, Iowa State University, Ames, Iowa
50011, USA}
\author{M.~A.~Tanatar}
\affiliation{Ames Laboratory, Ames, Iowa 50011, USA}
\affiliation{Department of Physics \& Astronomy, Iowa State University, Ames, Iowa
50011, USA}
\author{N.~D.~Zhigadlo}
\affiliation{CrystMat Company, CH-8037 Zurich, Switzerland}
\author{R.~Prozorov}

\affiliation{Ames Laboratory, Ames, Iowa 50011, USA}
\affiliation{Department of Physics \& Astronomy, Iowa State University, Ames, Iowa
50011, USA}
\date{\today}
\begin{abstract}
Anisotropic properties of single crystals of SrPt$_{3}$P were studied
using London penetration depth and electrical resistivity measurements.
The upper critical field, $H_{c2}(T)$, was determined from four-probe
electrical resistivity measurements for three orthogonal directions
of a magnetic field with respect to the crystal. The London penetration
depth, $\lambda(T)$, was determined from the magnetic susceptibility
of the Meissner-London state measured using a tunnel-diode resonator
technique. Whereas $H_{c2}(T)$ and the normal-state $\rho(T)$ are
practically identical for all three magnetic field orientations, the
London penetration depth shows significant unidirectional anisotropy.
The low-temperature $\lambda(T)$ is exponentially attenuated when
a small excitation radiofrequency magnetic field, $H_{rf}$, is applied along
the $c''-$direction, in which case screening currents flow in the
$a''b''-$ plane, while for the other two orientations, $H_{rf}\parallel a''$
and $H_{rf}\parallel b''$, the London penetration depth shows a much
weaker, $\lambda(T)\sim T^{2}$, variation. Such unusual and contrasting
behavior of the two anisotropies, $\gamma_{H}\left(T\right)=H_{c2,ab}/H_{c2,c}=\xi_{ab}/\xi_{c}$
and $\gamma_{\lambda}\left(T\right)=\lambda_{c}/\lambda_{ab}$, imposes
significant constraints on the possible order parameter. Although
our measurements are insufficient to derive conclusively the superconducting
gap anisotropy, qualitatively, order parameter with two point nodes
and a modulation in the perpendicular direction is consistent with
the experimental observations. 
\end{abstract}
\maketitle

\section{Introduction}

Superconductivity in platinum-based phosphides, APt$_{3}$P (A = Sr,
Ca, La), was discovered by Takayama \textit{et al.} in 2012 \cite{Takayama2012}.
Similar to non-centrosymmetric superconductors LaPt$_{3}$Si and CePt$_{3}$Si,
these compounds have distorted anti-perovskite structure but preserve
the inversion symmetry. The APt$_{3}$P compounds show apparently
large variation of the electron-phonon (EP) coupling strength from
weak to strong with the EP coupling constant values of, $\lambda^{EP}=$
0.57, 0.86 and 1.33, for LaPt$_{3}$P, CaPt$_{3}$P and SrPt$_{3}$P,
respectively \cite{Subedi2013,Aperis2020}.

SrPt$_{3}$P has superconducting transition temperature, $T_{c}\approx8.4$~K,
the highest among the 5d electron superconductors. From the heat capacity
measurements, Takayama \textit{et al.} concluded that SrPt$_{3}$P
is an $s-$wave superconductor. Its superconductivity is in a strong
coupling limit with the characteristic ratio, $\Delta\left(0\right)/k_{B}T_{c}\approx2.5$,
notably exceeding the weak-coupling value of $1.76$ \cite{Takayama2012}.
This feature was attributed to the presence of soft phonon modes.
The conclusion about the $s-$wave character of the superconducting
gap was further supported by the nuclear magnetic resonance (NMR)
measurements of the Knight shift, though no Hebel-Slichter peak was
observed \cite{Shiroka2015}. In an $s-$wave scenario, this feature
can also be due to the enhanced phonon damping. Nonlinear magnetic
field dependence of the Hall resistivity was interpreted as coming
from the multiple pockets of the Fermi surface, potentially hinting
at the multi-band superconductivity \cite{Takayama2012}. Hu \textit{et
al.} investigated the effect of Pd-doping in polycrystalline SrPt$_{3}$P
and found that there is a complex interplay between electron correlations,
electron-phonon coupling, and spin-orbit coupling \cite{Hu2016}.
In principle, such features may support unconventional order parameters
\cite{Ichioka2017,Kogan2011type15}.

To probe the nature of superconducting pairing in SrPt$_{3}$P, London
penetration depth was measured in polycrystalline samples using transverse-field
$\mu$SR \cite{Khasanov2014}. Combined with the measurements of the
critical field, the authors proposed that SrPt$_{3}$P is a two-band
superconductor with equal gaps but different coherence lengths in
different Fermi surface sheets. However, such a scenario is impossible
considering that there is only one characteristic length scale (one
$\xi$) for the spatial variation of the order parameter \cite{Kogan2011type15,Ichioka2017}.

On the theoretical side, from first principle calculations, Chen \textit{et
al.} suggested possible unusual superconductivity in SrPt$_{3}$P
caused by a charge density wave and strong spin-orbit coupling \cite{chen2012}.
In contrast, Subedi \textit{et al.} concluded that SrPt$_{3}$P is
a conventional $s-$wave superconductor in which spin-orbit coupling
plays only ``a marginal role" \cite{Subedi2013}. Many more theoretical
works studying the electronic structure and phonon modes followed
\cite{Nekrasov2012,Kang2013,Szczesniak2014,Zhang2016,Zocco2015,Jawdat2015}.

In this situation, it is important not only to establish the overall
thermodynamic behavior of SrPt$_{3}$P, but also to determine the
anisotropy of the superconducting order parameter. This cannot be
done on polycrystalline samples and requires crystals of sufficiently
large size in all directions. The growth of SrPt$_{3}$P single crystals
is non-trivial and requires high-pressure, high-temperature synthesis,
similar to MgB$_{2}$ \cite{Karpinski2007}. First single crystals
of SrPt$_{3}$P were synthesized in 2016 \cite{Zhigadlo2016}, and
they are used in this study.

In this paper, we report our investigation of the anisotropic London
penetration depth, $\lambda(T)$, the upper critical field, $H_{c2}(T)$,
and electrical resistivity, $\rho(T)$, in single crystals SrPt$_{3}$P.
Based on the results, we suggest that SrPt$_{3}$P may, indeed, support
an unconventional order parameter with point nodes.

\section{Experimental}

Single crystals of SrPt$_{3}$P were grown under high pressure in
a cubic anvil cell and were mechanically separated from the flux as
described in detail elsewhere \cite{Zhigadlo2016}. It has nearly
cubic crystal structure, $a=b=5.8$ \AA, $c=5.4$ \AA 
\cite{Zhigadlo2016}. Unfortunately, such difficult conditions result
in samples whose facets are not oriented in prime crystallographic
directions. We therefore assign three orthogonal axes based on sample
shapes: $a'$-, $b'$-, and $c'$-axes assigned for a resistivity
measurement sample, and $a''$-, $b''$-, and $c''$-axes for a London
penetration depth measurement sample. The shortest directions are
labeled as $c'$ and $c''$, the longest directions for $a'$ and
$a''$. Since the response in each of those directions is a linear
combination of the responses in ``true'' $a,b,c-$axes, observation
of electronic anisotropy would mean a true anisotropic response of
the crystal.

A crystal for the resistivity measurement was 0.6 mm long and 0.1$\times$0.1
mm$^{2}$ in cross-section. Silver wires of 50 $\mu$m diameter were
soldered to the sample using tin flux \cite{Tanatar2010a,Tanatar2011},
in a standard four-probe resistivity configuration. The contact resistance
was in the m$\Omega$ range. Resistivity measurements were performed
in a \textit{Quantum Design} PPMS down to 1.8~K in magnetic fields
up to 9~T in configurations $H\parallel a'-$axis, $H\parallel b'-$axis
and $H\parallel c'-$axis. Sample orientation was controlled by attaching
the sample to different sides of a plastic parallelopiped, see Ref.\cite{K122Ames2013}
for details. The upper critical field $H_{c2}$ was determined from
electrical resistivity measurements using different criteria as described
in the text below.

The temperature variations of the London penetration depth were measured
by using a self-oscillating tunnel diode resonator (TDR) when sample
temperature varied down to 400 mK ($\sim0.05~T_{c}$) \cite{VanDegrift1975RSI,Prozorov2000PRB}.
The TDR circuit resonates at 14 MHz and the frequency shift is measured
with one part per billion (ppb) precision. Its inductor coil generates
an ac magnetic field, $H_{rf}<20~\text{mOe}$, so that the sample
is always in the Meissner state at the temperatures of interest. The
size of the sample was 600 $\pm$ 5 $\mu$m ($a''-$axis), 250 $\mu$m
$\pm$ 5 $\mu$m ($b''-$axis), 155 $\pm$ 5 $\mu$m ($c''-$axis).
The sample was mounted on a 1 mm diameter sapphire rod and inserted
into an inductor coil of 2 mm diameter. The coil and the sample were
in vacuum in a $^{3}$He cryostat with sample and TDR circuit temperatures
actively stabilized by independent closed-loop LakeShore temperature
controllers. The London penetration depth was measured on the same
crystal in three different configurations; $H_{rf}\parallel a'',\:b''$,
and $c''-$axes. The resonant frequency of the $LC-$ tank circuit
containing the sample depends on the total inductance of the sample
in the coil, and it is straightforward to show that the frequency
shift compared to the empty resonator is proportional to the total
magnetic susceptibility of a sample up to a calibration factor that
depends on the sample dimensions, demagnetizing factor, and parameters
of the coil. The calibration constant is established for each experimental
run by mechanically pulling the sample out of the coil at the base
temperature of about 400 mK. Technical details of the technique are
provided elsewhere \cite{Prozorov2000PRB,Prozorov2000a,Prozorov2006SST,ProzorovKogan2011RPP}.

\section{Results}

\subsection{Upper critical field}

\begin{figure}[tb]
\centering \includegraphics[width=8.5cm]{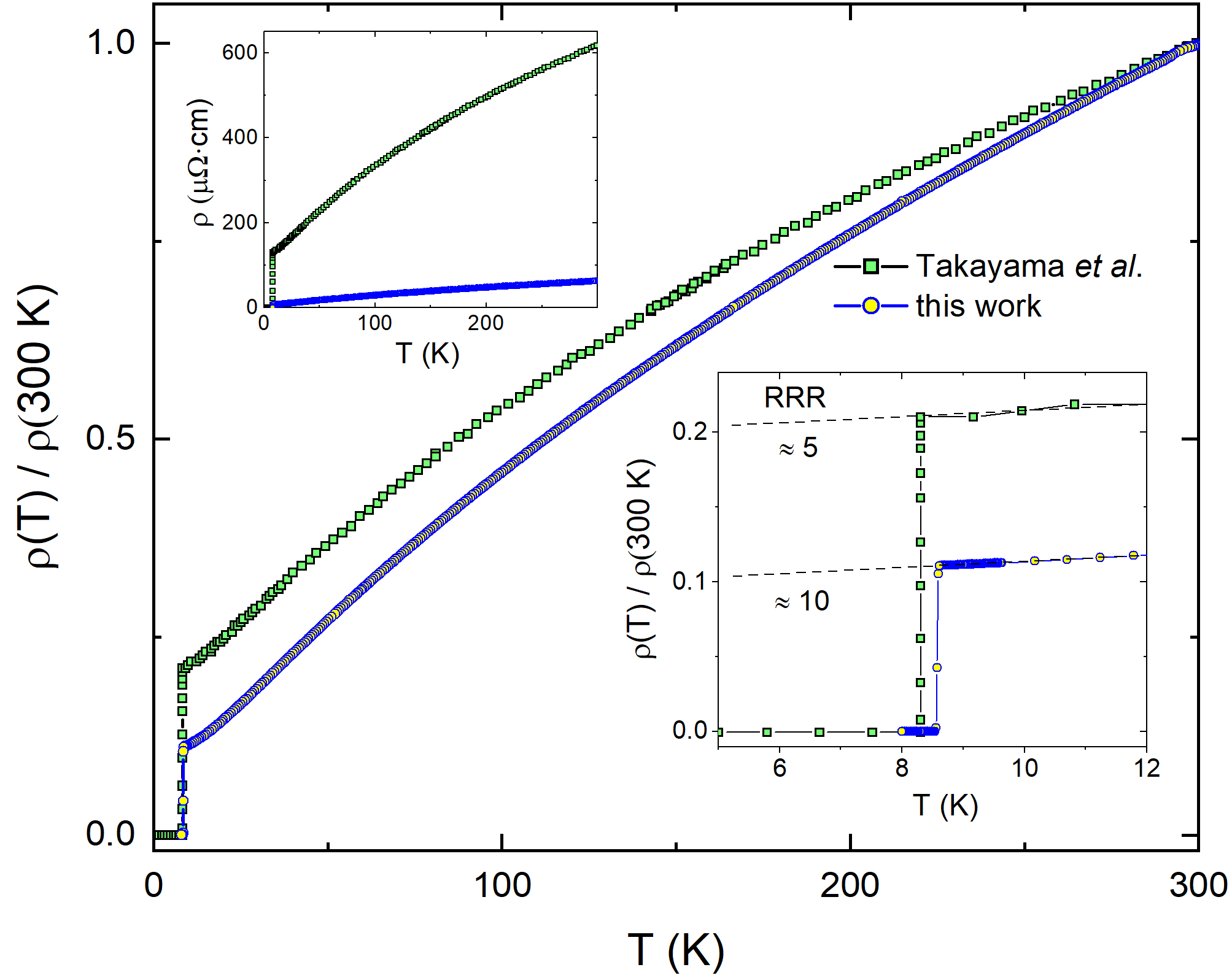} \caption{(Color online) Temperature dependent normalized resistivity in a single
crystal of SrPt$_{3}$P (blue circles) for current along the $a'-$axis.
It is compared to the results obtained in a polycrystalline sample
by Takayama \textit{et al.} (green squares) \cite{Takayama2012}.
The upper inset shows the non-normalized data, and the lower inset
zooms at the superconducting transition.}
\label{fig:R(T)} 
\end{figure}

Figure~\ref{fig:R(T)} shows the temperature-dependent resistivity
of SrPt$_{3}$P single crystal and compares it with the polycrystalline
sample data reported by Takayama \emph{et al.}\cite{Takayama2012}.
As shown in the upper-left inset, the single crystalline sample of
the present study has much lower resistivity over the whole temperature
range. The normalized resistivity in the main panel shows similar
temperature dependence between single crystal and polycrystalline
samples, but the residual resistivity ratio is higher in a single
crystal by a factor of two, RRR $\approx10$. The right-bottom inset
zooms at the superconducting transition. Note that the transition
temperature, $T_{c}$, is also somewhat higher in the single crystalline
sample. Assuming that a single crystal has a lower scattering rate
than a polycrystalline one, this implies a violation of the Anderson
theorem, which would be compatible with an anisotropic or nodal gap.

\begin{figure}[tb]
\centering \includegraphics[width=8.5cm]{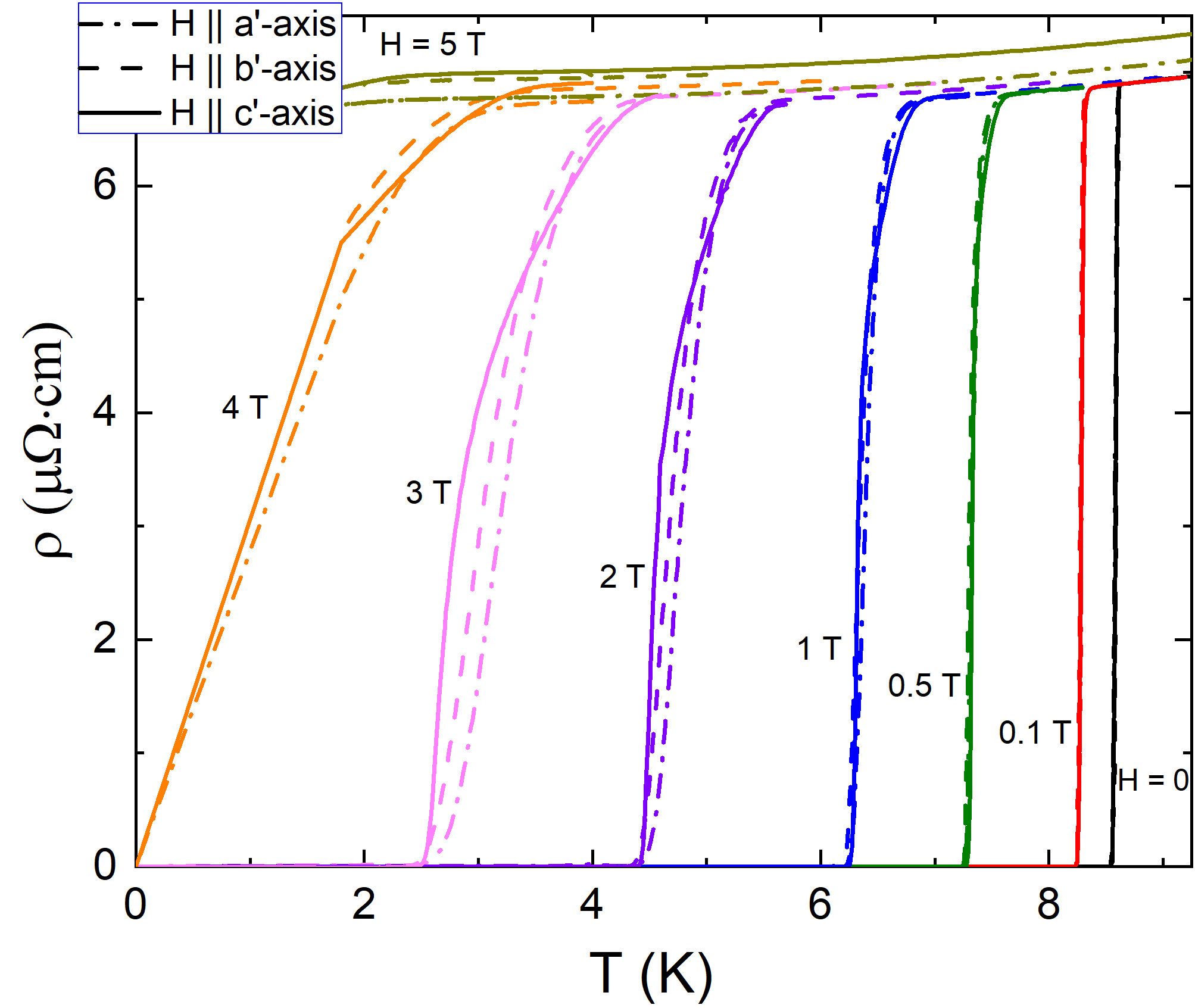} \caption{(Color online) Four-probe resistivity, $\rho\left(T\right)$, of SrPt$_{3}$P
single crystal measured in different magnetic fields (values indicated
next to the curves) in three orthogonal orientations, $H\parallel a'$
(dot-dashed), $b'$ (dashed), and $c'-$axes (solid).}
\label{fig:R(H)} 
\end{figure}

\begin{figure}[tb]
\centering \includegraphics[width=8.5cm]{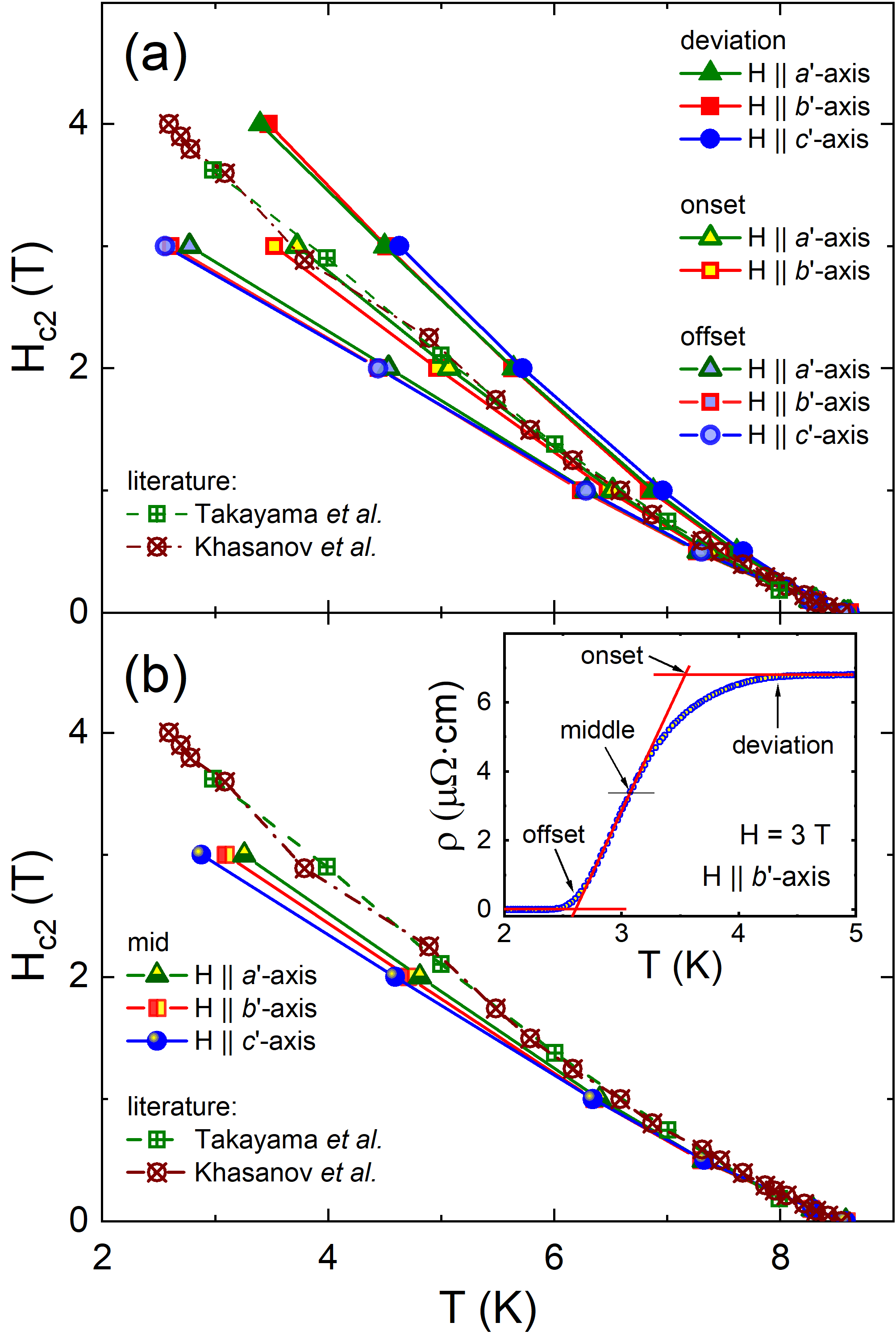} \caption{(Color online) Upper critical field, $H_{c2}(T)$ of SrPt$_{3}$P
single crystal measured with magnetic field applied along three principal
directions, $H\parallel a'-$axis (green symbols), $H\parallel b'-$axis
(red symbols) and $H\parallel c'-$axis (blue symbols). The inset
in panel (b) shows four different criteria used to define $H_{c2}$.
Upper panel (a) shows deviation (circles), onset (squares) and offset
(triangles) $H_{c2}$. Lower panel (b) shows commonly used midpoint-defined
$H_{c2}$. The results are compared with the data on polycrystalline
samples, from Takayama \textit{et al.} (crossed circles) \cite{Takayama2012}
and Khasanov \textit{et al.} (crossed squares) \cite{Khasanov2014}.}
\label{fig:Hc2} 
\end{figure}

The temperature-dependent resistivities, $\rho\left(T\right)$, measured
in the vicinity of the superconducting transition in magnetic fields
applied in three different orientations, $H$ $\parallel a',\:b'\:,\:c'-$axes,
respectively, are plotted in Fig.~\ref{fig:R(H)}. A very close to
isotropic behavior is self-evident. Perhaps, only $c'-$axis curves
become more rounded than the other two orientations in fields above
2 T. Fig.~\ref{fig:Hc2} summarizes the temperature-dependent upper
critical field, $H_{c2}(T)$, estimated using four different criteria
as shown in the inset in Fig.~\ref{fig:Hc2}(b). Top panel, Fig.~\ref{fig:Hc2}(a),
shows $H_{c2}(T)$ defined by the deviation, onset, and offset criteria.
Expectedly, the onset criterion produces values close to the literature
data on polycrystalline samples \cite{Takayama2012,Khasanov2014},
shown by open crossed symbols in both panels of Fig.~\ref{fig:Hc2}.
The commonly used mid-point transition data are plotted in Fig.~\ref{fig:Hc2}(b)
by different symbols and the same colors defined in panel (a). We
conclude that $H_{c2}(T)$ is very similar between all three orientations
regardless of the criterion used.

Returning to $H_{c2}\left(T\right)$, we note that often made assertion
that its upward (positive) curvature implies multi-band superconductivity
as claimed, for example, in Ref.\cite{Khasanov2014}, is not supported
by the theory \cite{Hc2ROPP2012,Hc2PRB2013,Anisotropies_PRB2019}.
Similar behavior, especially in the limited temperature interval,
may be the result of complications of the Fermi surface topology,
anisotropy of the order parameter, or nonmagnetic scattering even
in a single-band material \cite{Hc2PRB2013}. In fact, usually $H_{c2}(T)$
in multi-band superconductors shows ``normal'' concave behavior
and no difference in shape from the single-band result \cite{Hc2ROPP2012}.
Similarly, temperature-dependent anisotropy, $\gamma_{H}(T)=\xi_{ab}/\xi_{c}$,
does not imply multi-band superconductivity and is commonly found
in single-band superconductors in different circumstances \cite{Anisotropies_PRB2019}.

On the other hand, there may be consequences of multi-band character
on the response to scattering as discussed below. The important conclusion
from the resistivity measurements is that, as expected from the crystal
structure, this material is practically isotropic. Of course, the
electronic band structure is quite complicated, but for the analysis
of transport and thermodynamic properties, we may approximate it by
a Fermi sphere, an approach justified by the lack of anisotropy of
the upper critical fields.

\subsection{London penetration depth}

In a tetragonal system, London penetration depth has two components,
for fields penetration along the tetragonal plane, $\lambda_{a}$,
and for field penetration along the tetragonal axis, $\lambda_{c}$.
When a small magnetic field is applied in the plane, both $a-$ and
$c-$ components contribute to the measured signal, whereas for the
magnetic field along the $c-$axis, only the in-plane penetration
depth. In tetragonal SrPt$_{3}$P ($a=b\neq c$), therefore, intrinsically,
there are two distinct coherence lengths, $\xi_{a}=\xi_{b}\neq\xi_{c}$,
and two values of the London penetration depth, $\lambda_{a}=\lambda_{b}\neq\lambda_{c}$.
Since our crystals facets are not oriented properly along the prime
crystallographic directions, we use $a'-, b'-, c'-$ axes for one crystal, and
$a''-, b''-, c''-$ axes for another. However, the resultant measured quantities
are the linear combinations of the crystallographic quantities, therefore
we have two distinct characteristic lengths of both types, $\xi$
and $\lambda$.

We now examine the temperature variation of the London penetration
depth, $\lambda(T)$, which is linked directly to the structure of
superconducting gap as it depends sensitively on the thermally excited
quasiparticles. High-resolution magnetic susceptibility was measured
at zero applied dc field in three different orientations of small
excitation field, $H_{rf}<20$~mOe, along $a''-$, $b''-$ and $c''-$axes
as discussed in the experimental section above.

\begin{figure}[tb]
\centering \includegraphics[width=8.5cm]{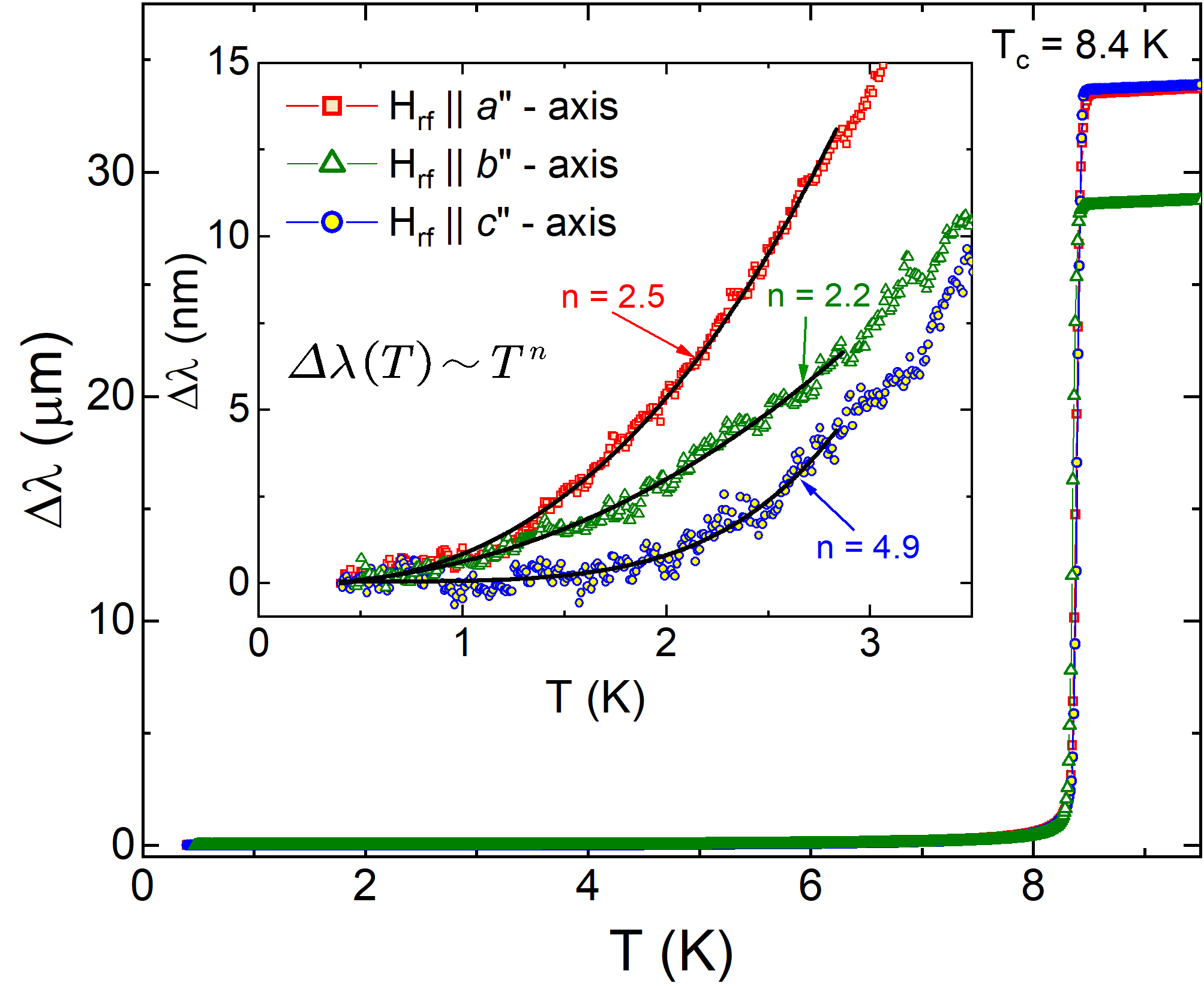} \caption{(color online) Temperature dependence of the London penetration depth
measured for three orthogonal orientations: $\lambda_{H_{rf}\parallel a''}(T)$
(red squares), $\lambda_{H_{rf}\parallel b''}(T)$ (green triangles),
and $\lambda_{H_{rf}\parallel c''}(T)$ (blue circles). Inset zooms
at the low-temperature region of the same curves, green and blue symbols,
respectively. The black solid lines are the fits in the range of $T\protect\leq T_{c}$/3,
to the power-law: $\Delta\lambda\sim T^{n}$.}
\label{fig:penetration} 
\end{figure}

Figure~\ref{fig:penetration} shows the low-temperature variation
of the London penetration depth; $\lambda_{H_{rf}\parallel a''}(T)$
(red squares), $\lambda_{H_{rf}\parallel b''}(T)$ (green triangles),
and $\lambda_{H_{rf}\parallel c''}(T)$ (blue circles). As clearly
shown in the inset, the low-temperature part of the London penetration
depth exhibits distinct behavior depending on the magnetic field orientation.
London penetration depth depends on the integral of the projection
of the Fermi velocity over the whole Fermi surface and having nodes
or anisotropy in some parts of the order parameter has an effect on
all components of $\lambda$ \cite{Prozorov2006SST,ProzorovKogan2011RPP}.
In addition, scattering may play a significant role and, depending
on the order parameter structure, even non-magnetic (potential) scattering
can be pair-breaking, which significantly affects the penetration
depth and effectively turns exponential into a power-law behavior
at low temperatures \cite{Prozorov2006SST,ProzorovKogan2011RPP}.
Therefore, we use a power-law fitting, $\Delta\lambda\sim T^{n}$,
to quantify the degree of creation of quasiparticles with increasing
temperature in different directions. The inset in Fig.~\ref{fig:penetration}
shows the large exponent, $n=4.9$, for $\lambda_{H_{rf}\parallel c''}(T)$.
Numerically, the exponents above $n=4$ are indistinguishable from
exponential and indicate a fully gapped superconducting order parameter.
On the other hand, the lower exponents were obtained for $\lambda_{H_{rf}\parallel a''}$$(T)$
($n=2.5$) and $\lambda_{H_{rf}\parallel b''}$$(T)$ ($n=2.2$),
signifying either line-nodal gap in the dirty limit, the multi-band
sign-changing order parameter, such as $s_{\pm}$ in the dirty limit,
or a point node \cite{Prozorov2006SST,ProzorovKogan2011RPP}. However,
if the order parameter had a line node or was an $s_{\pm}$ type on
a three-dimensional Fermi surface, the measurements for all three
orientations would show similar low power $n$. Our data seem to suggest
a case with the point node.

\begin{figure}[tb]
\centering \includegraphics[width=8.5cm]{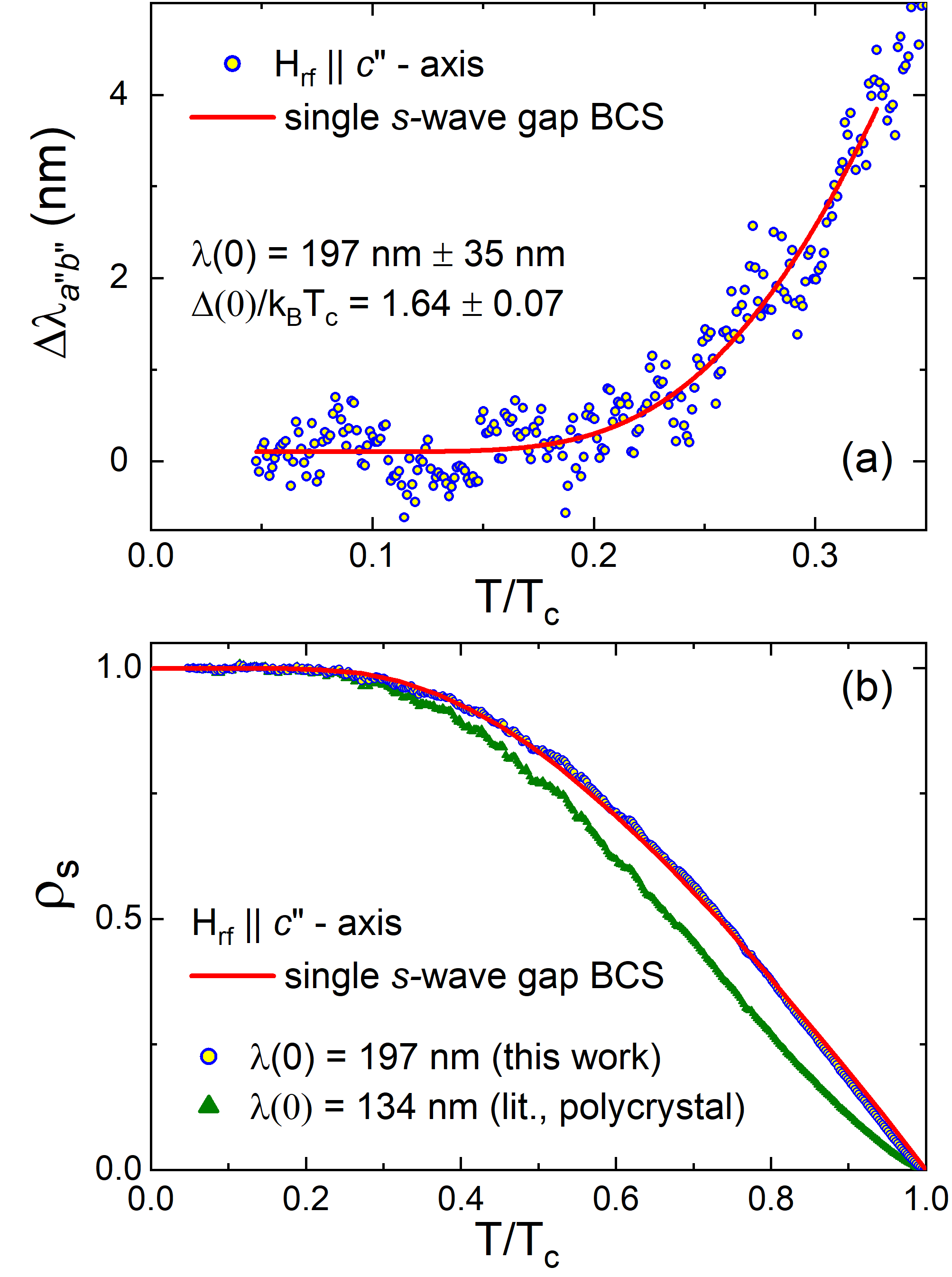} \caption{(color online) (a) Low-temperature part of the London penetration depth for ${H_{rf}}\parallel c''$. The red solid line is a single s-wave gap BCS fit. (b) Normalized superfluid density of the same data in the panel (a), $\rho_{s}(T)=\left(\lambda(0)/\lambda(T)\right)^{2}$, plotted with  $\lambda_{0}$
= 197 nm (blue circle, this work) and $\lambda_{0}$
= 134 nm (green triangle, from a polycrystal, Ref~\onlinecite{Khasanov2014}).}
\label{fig:superfluid} 
\end{figure}

For the data of high exponent $n$ = 4.9 ($\lambda_{H_{rf}}\parallel c''$),
we attempted a BCS fitting with a standard expression, $\Delta\lambda(T)=\lambda(0)\sqrt{\pi\delta(0)/2T}\exp(-\delta(0)/T)$.
The best fitting is shown in Fig~\ref{fig:superfluid} (a) with $\lambda_{0}$
= 197 nm $\pm$ 35 nm. In the panel (b), its superfluid density, $\rho_{s}(T)=\left(\lambda(0)/\lambda(T)\right)^{2}$,
was plotted which is consistent with a single gap BCS case. For comparison,
the superfluid density was also plotted with $\lambda_{0}$
= 134 nm obtained from a polycrystalline sample (ref.~\onlinecite{Khasanov2014}).

\section{Discussion}

\subsection{The relations between different anisotropies}

In tetragonal crystal with $a=b\neq c$, the anisotropies of the upper
critical field, $\gamma_{H}=H_{c2,ab}/H_{c2,c}=\xi_{ab}/\xi_{c}$,
and of the London penetration depth, $\gamma_{\lambda}=\lambda_{c}/\lambda_{ab}$,
depend on the the anisotropy of the Fermi surface and of the order
parameter \cite{Hc2ROPP2012,Kogan2020,Anisotropies_PRB2019}. Here,
$H_{c2,c}=\phi_{0}/2\pi\xi_{ab}^{2}$ is the upper critical field
measured with a magnetic field applied along the $c-$axis and $\xi_{ab}$
is the coherence length in the transverse, in-plane direction, $H_{c2,ab}=\phi_{0}/2\pi\xi_{c}\xi_{ab}$
is the in-plane upper critical field; $\lambda_{c}$ is the London
penetration depth of the in-plane magnetic field with screening super-currents
flowing along the $c-$axis, and $\lambda_{ab}$ is for the magnetic
field along the $c-$axis and screening currents in the $ab-$plane.
In the text we use index $a$ with the understanding that it is equivalent
to $b$, hence $ab$. These anisotropies are generally temperature-dependent
and may increase or decrease on warming or cooling, with $\gamma_{H}\left(T\right)$
and $\gamma_{\lambda}\left(T\right)$ often going in the opposite
directions \cite{Anisotropies_PRB2019,Kogan2020}. This is easier
to see if we assume a commonly used separation of variables ansatz,
$\Delta\left(T,\mathbf{k}\right)=\Psi\left(T\right)\Omega\left(\mathbf{k}\right)$,
with the normalization $\left\langle \Omega^{2}\right\rangle _{FS}=1$,
where $\left\langle ...\right\rangle _{FS}$ denotes the averaging
over the Fermi surface. For example, isotropic $s-$wave is described
by $\Omega=1$, whereas a $d-$wave order parameter is given by $\Omega=\sqrt{2}\cos2\varphi$.
At $T_{c}$, in clean case, we always have $\gamma_{H}^{2}\left(T_{c}\right)=\gamma_{\lambda}^{2}\left(T_{c}\right)=\left\langle \Omega^{2}v_{a}^{2}\right\rangle /\left\langle \Omega^{2}v_{c}^{2}\right\rangle $,
so the anisotropy depends on the symmetry of the order parameter described
by $\Omega\left(\mathbf{k}\right)$ \cite{Anisotropies_PRB2019}.
The electrical conductivity is, $\sigma_{ik}=2e^{2}N\left(0\right)D_{ik},$
where $D_{ik}=v_{i}v_{k}\tau$ is the diffusivity tensor. In these
equations, $v_{i}$ is the relevant component of the Fermi velocity,
$N\left(0\right)$ is the density of states at the Fermi level, and
$\tau$ is the transport scattering time of the normal metal. Therefore,
the anisotropy of resistivity is, roughly, $\gamma_{\rho}\left(T_{c}\right)=\left\langle v_{a}^{2}\right\rangle /\left\langle v_{c}^{2}\right\rangle $
assuming isotropic $\tau$. Therefore, for isotropic order parameter,
$\Omega=1$, regardless of the Fermi surface anisotropy, $\gamma_{H}^{2}\left(T_{c}\right)=\gamma_{\lambda}^{2}\left(T_{c}\right)=\gamma_{\rho}\left(T_{c}\right)$.
However, in general, $\gamma_{H}^{2}\left(T_{c}\right)=\gamma_{\lambda}^{2}\left(T_{c}\right)\neq\gamma_{\rho}\left(T_{c}\right).$
There is one more possibility. In the dirty limit, the anisotropy
of the order parameter washes away, and the equality of all three
$\gamma$ is restored. However, there is no reason to assume that
we are at the dirty limit.

Electronic bandstructure calculations predict isotropic electrical
resistivity of the normal state \cite{Kang2013}, and we observed
it in our experiments. Combined with the measured nearly isotropic
upper critical field at $T_{c}$, this would imply isotropic $\Omega=1$
or a dirty limit. However, significantly anisotropic London penetration
depth represents a problem for this interpretation. In particular,
its nearly exponential attenuation at the low temperature when the
excitation magnetic field, $H_{rf}$, is applied along the $c''$
axis, but substantially lower exponents along the other two orthogonal
directions, not far from $n=2$, when the power-law fitting, $\lambda=AT^{n}$,
is performed.

In our view, the only way to understand such behavior within a standard
theory of superconductivity without invoking some complicated scenarios
is to find an order parameter for which $\gamma_{H}^{2}\left(T_{c}\right)=\gamma_{\lambda}^{2}\left(T_{c}\right)=\gamma_{\rho}\left(T_{c}\right)=1.$
Importantly, it does not imply that $\gamma_{\lambda}\left(T\right)$
is isotropic at the intermediate temperatures. Several order parameters
were previously analyzed for their resulting anisotropies \cite{Anisotropies_PRB2019,Kogan2020},
and some of them satisfy the requirement.

\begin{figure}[tbh]
\centering \includegraphics[width=8.5cm]{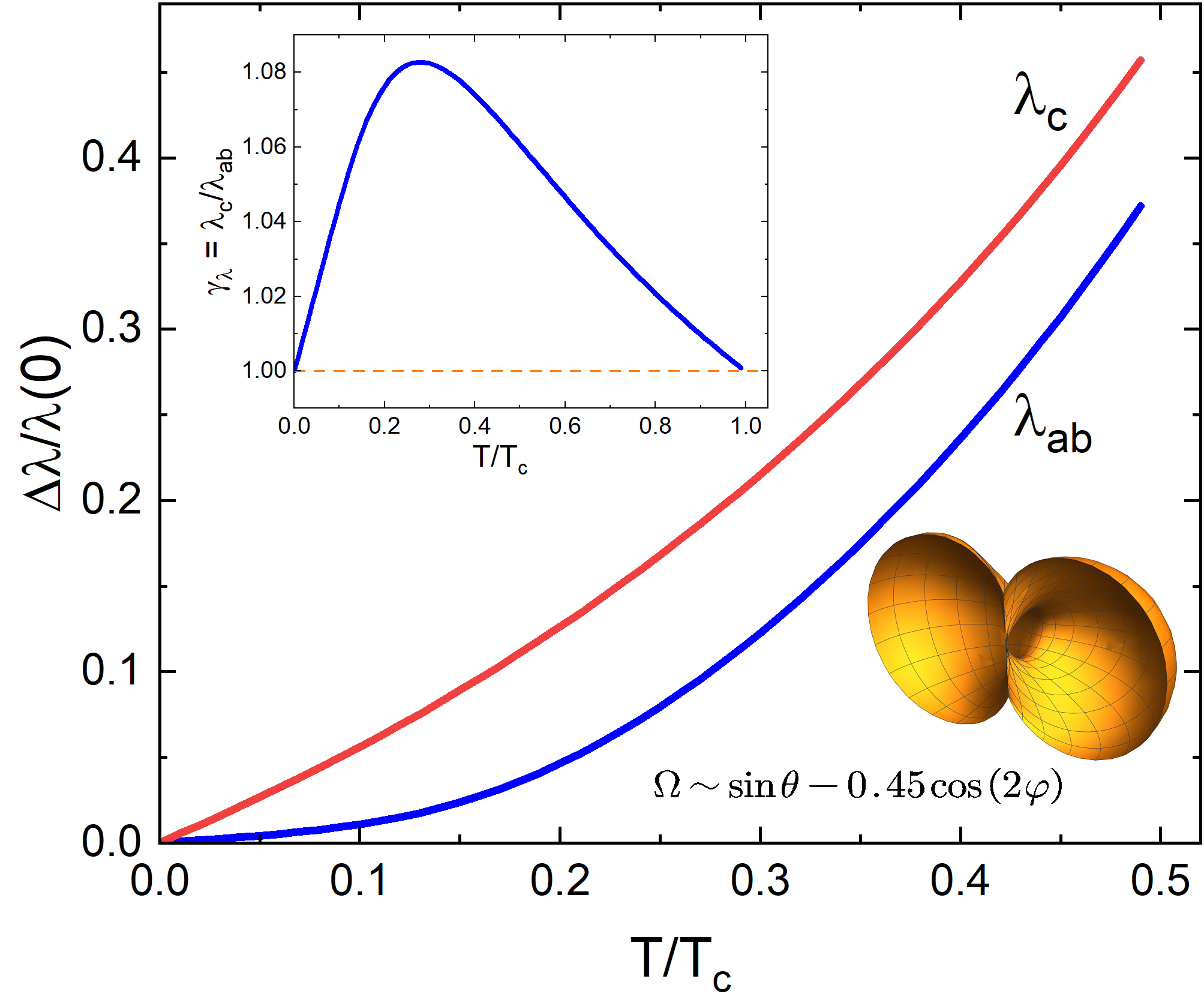} \caption{(Color online) The variation of the London penetration depth in two
orientations for the order parameter, $\Omega/\Omega_{0}=\sin\theta-0.4529\cos2\varphi$,
where the numeric coefficient was chosen so that the system is isotropic
at $T=0$ and $T_{c}$. This order parameter, with two polar point
nodes, in the $k-$ space is shown in the inset.}
\label{theory} 
\end{figure}

Of course, we do not know the type of the order parameter, and our
crystals are not properly oriented, so this is just one plausible
order parameter fully compatible with our observations. Specifically
we consider a composite order parameter, $\Omega\propto\sin\theta-a\cos2\varphi$,
which is a combination of a $d-$wave - like variation in the $ab-$plane
(azimuthal angle $\varphi$) and having two point nodes at the poles,
at the polar angle, $\theta=0,\pi$, as shown schematically in Fig.\ref{theory}.
Point nodes are known to result in a power-law behavior for out-of-plane penetration depth, whereas showing nearly exponential variation
in the plane \cite{Einzel1986,Kogan2020}. The reason that we chose
this more complicated form of the order parameter, and not just $\Omega\propto\sin\theta$
is that at $a=0.4529$, it produces a non-monotonic $\gamma_{\lambda}\left(T\right)$
shown in the inset in Fig.\ref{theory}, that starts at $T=0$, and
ends at $T_{c}$ at the isotropic value of 1. As shown in Fig.\ref{theory},
this order parameter results in a significant difference in $\lambda\left(T\right)$
between different orientations exhibiting behavior similar to our
observations.

Our observations of presumably point-node superconducting gap structure should be contrasted with the chiral topological superconducting state observed in polycrystalline samples of the closely related LaPt$_3$P \cite{Biswas2021}. The two materials are isostructural but differ by the electron count. In the LaPt$_3$P the temperature-dependent superfluid density is consistent with nodal gap, potentially with horizontal line node and polar point nodes, as determined from $\mu$SR measurements. Changing electron count moves the system to clearly point-node behavior, with line nodes erased. This observation clearly suggests that Fermi surface topology change plays important role in the superconducting pairing of the compounds. Further investigations may be of great importance. 

\section{Conclusions}

Based on the measurements of the upper critical fields, the London
penetration depths, and their anisotropies, our results open up a
possibility of a nontrivial order parameter in SrPt$_{3}$P, possibly
with point nodes somewhere on the isotropic Fermi surface. Taking
into account considerable difficulty of making single crystals of
this compound, this conclusion is a step forward compared to the polycrystalline
samples where such observation would be impossible. We hope our work
will motivate further effort to grow and measure better single crystals
of SrPt$_{3}$P with well-defined orientations of the facets. 
\begin{acknowledgments}
We thank Seongyoung Kong and Kirill Kovnir for help with crystallographic orientation, and Linlin Wang for useful discussions of the
electronic structure. Work in Ames was supported by the U.S. Department
of Energy (DOE), Office of Science, Basic Energy Sciences, Materials
Science and Engineering Division. Ames Laboratory is operated for
the U.S. DOE by Iowa State University under contract DE-AC02-07CH11358.
N.D.Z. acknowledges support from the Laboratory for Solid State Physics
ETH Zurich and the Department of Chemistry and Biochemistry of the
University of Bern where crystal growth studies were initiated. 
\end{acknowledgments}

\section{Data Availability}
All data that support the findings of this study are included within the article (and any supplementary information files).

\bibliographystyle{apsrev4-2}
\bibliography{SrPt3P}

\end{document}